\documentclass[12pt,english]{article}
\usepackage{graphicx}
\usepackage{epsfig}
\usepackage[english]{babel}
\usepackage[ansinew]{inputenc}
\usepackage[T1]{fontenc}
\usepackage{latexsym}
\usepackage{amssymb}

\begin{document}
\title{
%\hfill {\large Internal notes}\\
%\hfill {\normalsize Neutrinica-2009-13}\\
%\hfill {\normalsize draft --- not for distribution}\\
Simulations of neutrino oscillations for a low-energy neutrino factory with a magnetized large-volume liquid scintillator at 2288 km baseline}
\author{Juha Peltoniemi\\ 
{\em Neutrinica Ltd, Oulu, Finland}\\ and \\
\em Excellence Cluster Universe \\ \em Technische Universität München, Garching, Germany}
\date{\today}
\maketitle

\begin{abstract}
A magnetized large-volume liquid scintillator might be used as a tracking detector with charge identification to measure high-energy neutrinos. I consider neutrino oscillations for a beam from a low-energy neutrino factory to such a detector for the baseline CERN-Pyh\"asalmi, with the length of 2288 km. The range to study the oscillation parameters extends to $\sin^2 2\theta_{13}\sim 10^{-4}$ for reasonable assumptions for the detector and the beam.
\end{abstract}

\section{Introduction}

I study the low-energy neutrino factory \cite{Ankenbrandt:2009zz, Long:2009zz, Bross:2009gk, Geer:2007kn, Bandyopadhyay:2007kx, Abe:2007bi, :2008xx} with the 2288 km long baseline CERN-Pyh\"asalmi, using a magnetized large volume  liquid scintillator. As the prototype model I take a magnetized LENA\cite{Wurm:2007cy, Oberauer:2006cd, Hochmuth:2006gz, MarrodanUndagoitia:2006qs, MarrodanUndagoitia:2006qn, MarrodanUndagoitia:2006rf, MarrodanUndagoitia:2006re, Undagoitia:2005uu, Autiero:2007zj}, a 30 m wide cylinder in vertical orientation, by default with 50 kton fiducial mass. 

It has been suggested that a large volume liquid scintillator indeed has capacity to measure high-energy neutrinos \cite{Peltoniemi:2009xx, Learned:2009rv}. On-going studies hint that with a good photodetection a magnetic liquid scintillator can be used to measure the charge of muons. The performance may depend on the design of the detector and on the choices of the components. Here I assume an optimistic performance and study different options, to set the goals for the design of the experiment.

Previously a conventional beam of 1--6 GeV from CERN to LENA at Pyh\"asalmi mine has been considered\cite{Peltoniemi:2009hv}, as well as high-energy beta beams \cite{Peltoniemi:2009zk}. The conventional beam is a very viable option to study neutrino parameters if the third mixing angle is not too small, i.e. $\sin^2 2\theta_{13} >3\cdot10^{-3}$, and the beta beam with the same setup might extend the range to $\sin^2 2\theta_{13} >1\cdot10^{-3}$. With a larger detector or stronger beams those limits might be extended by a factor of 2--5. 

The Pyh\"asalmi Mine provides a 1400 m (4000 m.w.e.) deep underground site in the middle of Finland. The density profile for the CERN-Pyh\"asalmi baseline has been modelled well \cite{Kozlovskaya:2003kk,Peltoniemi:2006hf}, and the accuracy of the average density can be taken 1 \%. In these simulations, a simplified profile is used \cite{prem}.

\section{Low-energy neutrino factory}

\begin{figure}[tbp]
\begin{center}
\includegraphics[angle=0, width=10cm]{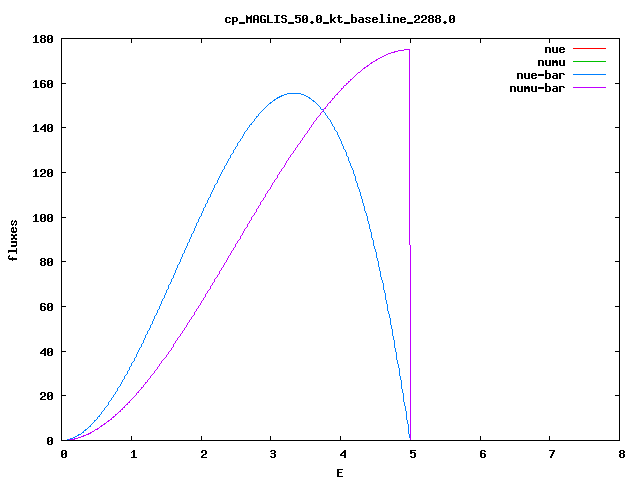}
\caption{\sf Spectrum of neutrinos from a 5 GeV muon beam. }
\label{betaspectrum}
\end{center}
\end{figure}

For the CERN-Pyh\"asalmi baseline the first oscillation peak is at about 4.2 GeV with usual parameters. Hence a beam with an energy range 1--5 GeV appears the most interesting for this baseline. 

Here I study neutrinos made in a neutrino factory by the decay of high-energy muons: $\mu^- \to \nu_\mu e^- \overline{\nu}_e$, and respectively for the positive charge. The golden signal is the wrong-sign muon appearance, as well as the respective disappearance channels. A wrong-sign electron appearance would be more challenging (platinum channel). The tau appearance (silver channel) is relevant for higher energies. 

For the detector a technology to measure the neutrino helicity is required. In practise the lepton charge is measured using magnetic field. Non-magnetized detectors with a low-energy neutrino factory were studied in \cite{Huber:2008yx}. 

The technology for the neutrino factory is currently under development. It is not yet completely clear how feasible it will be, but in any case the low energy neutrino factory (5 GeV) is expected to be substantially cheaper than the ultimate high-energy neutrino factory (20--50 GeV). 

I assume for the beam power $1.1\cdot10^{21}$ usable muon decays per year, and 5 year running time for both polarities.

The detector capable to measure neutrinos from the neutrino factory will also work well for conventional or beta beams. While the beta beam may be rather a competitor for the neutrino factory, a conventional beam could be considered complimentary, or rather a predecessor to the neutrino factory, given that the cost of the beamline will be a tiny fraction the cost of the neutrino factory. Probably the neutrino factory would be built only after the preceding beam fails to measure the interesting quantities due to $\theta_{13}$ being out of range. Hence I assume here that there will be an accompanying conventional wide band beam, as described in \cite{Peltoniemi:2009hv}.

\section{Detector}

As the detector model I take a magnetized LENA. The plain LENA is planned to consist of 50 kton of liquid scintillator, in a vertical cylindrical tank with the radius of 12 m and the height of 100 m. It is surrounded by a buffer of 2 m, and outside it is shielded by water (totally 100 kton). The scintillation light is read by more than 10 000 photosensors.

I assume that the detector can be magnetized, either surrounding the tank with a circular coil to generate a homogeneous vertical field inside the tank (and outside a inhomogeneous cylidnrically symmetric field in the opposite direction) or otherwise.

With a good photodetection system the large volume liquid scintillator can be used as a tracking detector. One or more tracks can be reconstructed from the measured photon time profiles. With a good reconstruction one can define the lepton flavor and estimate the neutrino energy. This requires a fast scintillator (3 ns) and a good time resolution for the photosensors (better than 3 ns), as well as an ability to count multiple photons arriving in a single photosensor with a reasonable accuracy. While the considered concept is formally a totally active scintillator detector, it is fundamentally different from the segmented scintillator \cite{Abe:2007bi} which the term TASD usually refers to.

\begin{figure}[tbhp]
\begin{center}
\includegraphics[angle=0, width=15 cm]{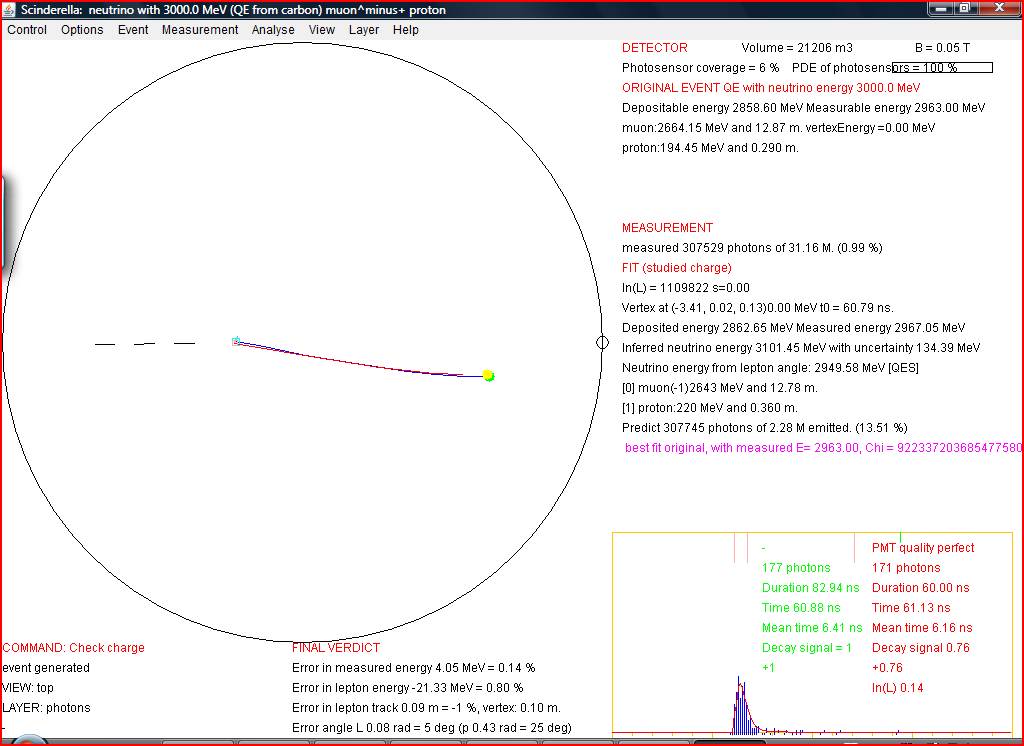}
\caption{\sf A screenshot from a simplified simulation program, with $B=0.05$ T, showing how to resolve the curved track for a multi-GeV muon using the photon observation time information of all individual photosensors, with an optimal photodetection system. The blue line is the original muon track and the red one the fit. }
\label{fit}
\end{center}
\end{figure}

The lepton charge can be defined if its curvature under magnetic field is measurable. Initial studies point that the required magnetic field to cause an observable bending is about 0.02--0.05 T, depending on the time properties of the scintillator and the photosensors, and also on the event topology (See Fig.~\ref{fit}). On the other hand, the magnetic field must be strong enough to separate the magnetic curvature from stochastic deviations, mostly multiple Coulombian scattering. The scale of the multiple scattering is defined by the scattering length, in liquid scintillator 42 cm, compared with 1.8 cm in iron. For typical track lengths the magnetic bending is equal to the mean stochastic deviation with $B=0.02$ T, and to get sufficient statistical confidence at least $B=0.1$ T is required. Further simulations are required to define the optimal magnetic field to reach a sufficiently low misidentification probability, particularly because the multiple scattering is not Gaussian.

The magnetization requires photosensors that work under such a magnetic field. The availability of such detectors of sufficient time resolution and dynamic range is not yet clear, but we have several years to develop such devices before the decisions for the neutrino factory are due. Decisions to construct LENA may be, however, due earlier than that, and for that the selection of the photosensor technology is a rather urgent task, as well as the decision to magnetize the detector (or at least to keep an option for it).

Other than the magnetization, there are some chances to distinguish the neutrino helicity by other methods, though only statistically\cite{Huber:2008yx}. The most significant may be identifying the recoil nucleon --- the identification is rather clear but due to possible intranuclear interactions the nucleon flavor may be changed at 10--20 \% probability, and the presense of charged pions with unidentified charge spoils the possibility altogether. Given the relatively moderate cost of the magnetization compared with the total cost of the experiment, using a non-magnetized scintillator may not be a cost-effective way to go, unless the magnetization turns out to be unfeasible due to unavailability of suitable photosensors or other reasons. For the platinum channel (electron appearance), however, the given statistical methods may be the only way to get charge information unless the photodetection can be improved beyond present assumptions (e.g. by photographic methods).   
 
I consider also a larger size, 150 kton, for comparison. This could be achieved by enlarging the size, using the outer shield as additional volume or building another dedicated beam detector, probably aligned towards the beam. Increasing the beam power (which is a very unknown quantity) would have a similar effect as increasing the detector size. Given that the large-volume liquid scintillator is expected to be much cheaper than a segmented scintillator or iron calorimeter (and a dedicated high-energy detector costs only a fraction of an all-range detector), in this case larger detectors are more affordable and even desirable to balance the total investment. Anyway, the detector will have also a research programme of its own, for atmospheric neutrinos, proton decay, cosmic rays and low-energy neutrinos.

Throughout this work I assume the energy resolution of 5 \%, unless otherwise stated. The resolution depends quite substantially on the properties of the scintillator, phototubes and electronics, but according to on-going work, this assumption is realistic, unless the photoaccuracy is spoiled by the magnetisation. From the analysis point, the magnetization itself just improves the accuracy of energy definition. 

The detection aspects will be described in more detail elsewhere. 

\section{Simulations}

The simulations are done using the GLOBES codebase\cite{Huber:2007ji,Huber:2004ka}, embedded within an own code. The AEDL file described in \cite{huber:2002mx} was used as a starting point.

\begin{figure}[tbhp]
\begin{center}
\includegraphics[angle=0, width=7cm]{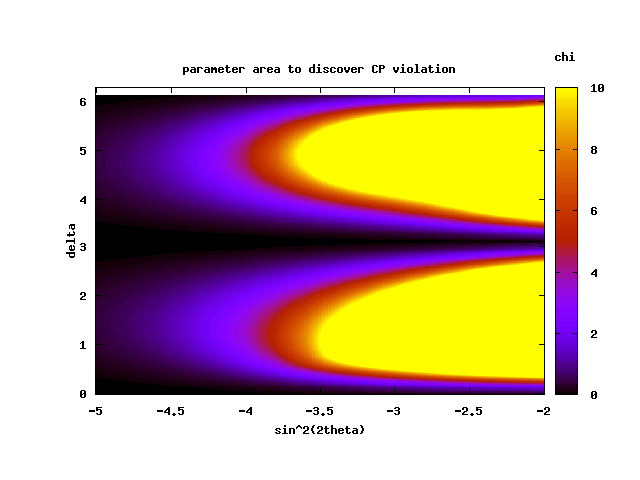}
\includegraphics[angle=0, width=7cm]{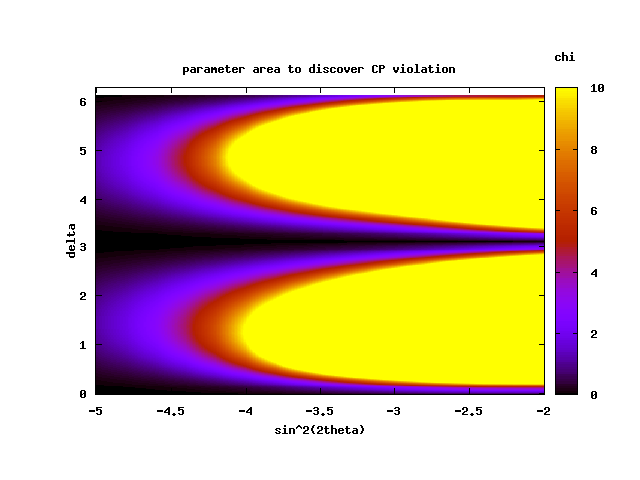}
\caption{\sf The CP range with a combined set of a neutrino factory and a wide band beam for the same 2288 km baseline. Left 50 kton and right 150 kton. There is a major difference, the larger is much better.
The colors represent $\chi$ values, so that yellow is about $3\sigma$, red $2\sigma$ and blue $1\sigma$.}
\label{cp_nw}
\end{center}
\end{figure}

The hottest motivation for the long baseline oscillation experiments is to discover leptonic CP violation. This is expressed with the parameter $\delta$, values differing from 0 or $\pi$ meaning CP violation.
The observable range of the CP angle $\delta$ for the assumed beam and baseline is shown in Fig.~\ref{cp_nw}, with both 50 kton and 150 kton fiducial masses. At best the range extends to $\sin^2 2\theta_{13} \sim 8\cdot10^{-5}$ with 150 kton, with 50 kton $\sin^2 2\theta_{13} \sim 2\cdot 10^{-4}$. This is significantly better than that achievable with conventional beams or beta beams, though a fair comparison depends on the achievable beam powers.

\begin{figure}[tbhp]
\begin{center}
\includegraphics[angle=0, width=7cm]{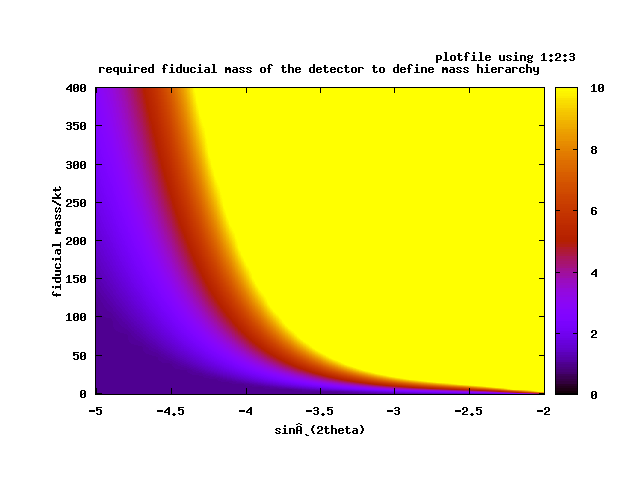}
\includegraphics[angle=0, width=7cm]{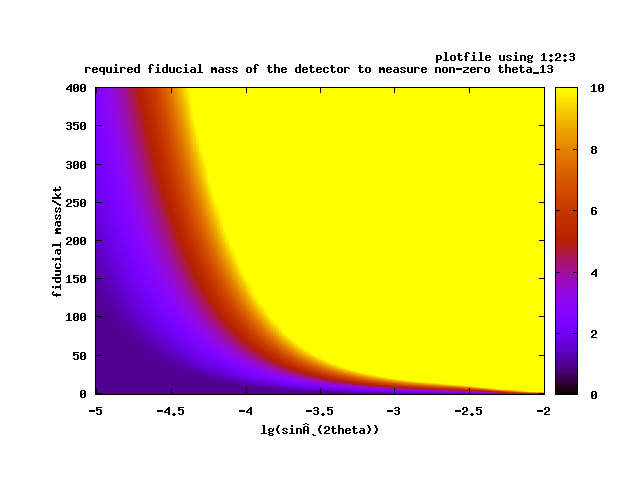}
\caption{\sf The capacity to measure the mass hierarchy (left) and the $\theta_{13}$ reach (right), with both a neutrino factory and a wide band beam together. In the horizontal axis is the fiducial mass of the detector, used as a scaling factor including also variations in the beam power, the detector efficiency and the running time. Here and in the other similar figures $\delta=3\pi/2$. }
\label{mhf_bw}
\end{center}
\end{figure}

The considered baseline is particularly suitable to study the mass hierarchy, i.e. the sign of $\delta m_{23}$. 
The neutrino factory extends the search of the mass hierarchy to smaller mixing angles than any other beam. In the given examples, with 150 kton detector $\sin^2 2\theta_{13} \sim 3\cdot 10^{-4}$ is reachable for the worst $\delta$, and for the best case $\sin^2 2\theta_{13} \sim  8\cdot10^{-5}$. Without the assumed conventional beam the performance curve at $\delta$ plane has glitches towards lower miging angles, i.e. the conventional beam improves the performance at lower mixing angles. 

\begin{figure}[tbhp]
\begin{center}
\includegraphics[angle=0, width=7cm]{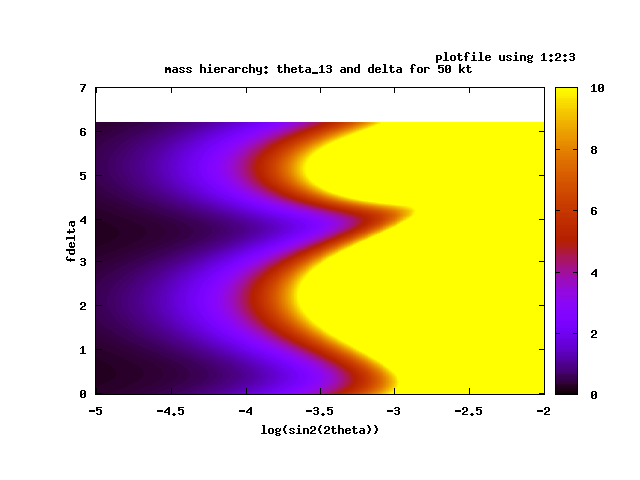}
\includegraphics[angle=0, width=7cm]{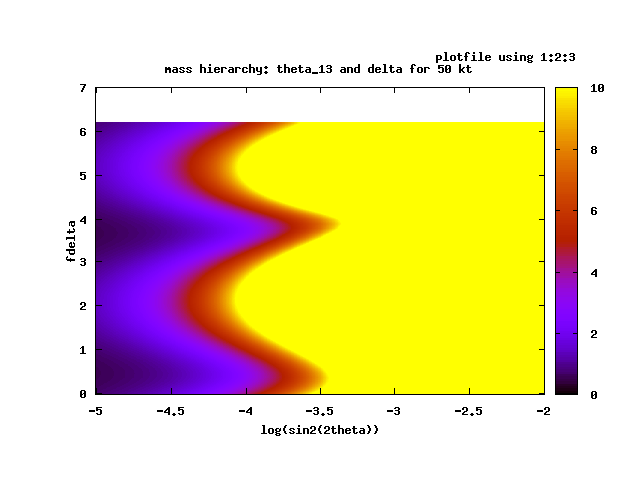}
\caption{\sf The capacity to measure the mass hierarchy in the plane of $\theta_{13}$ and $\delta$, with 50 kton and 150 kton detectors and neutrino factory and conventional beam together.}
\label{mh_bw}
\end{center}
\end{figure}

Measuring the last unknown mixing angle, $\theta_{13}$ is also one of the main goals of the experiment. If other experiments measure it before neutrino factory, the neutrino factory will be rather a precision machine than discovery machine (though measuring a large mixing might detract seriously from the motivation to build the neutrino factory).
The range to measure $\theta_{13}$ extends to $\sin^2 2\theta_{13} \sim (2\dots 9)\cdot 10^{-4}$ with 50 kton, or $\sin^2 2\theta_{13} \sim (0.8\dots 3)\cdot 10^{-4}$ with 150 kton. Without the conventional beam the reach with the worst $\delta$ is poorer.

\begin{figure}[tbhp]
\begin{center}
\includegraphics[angle=0, width=7cm]{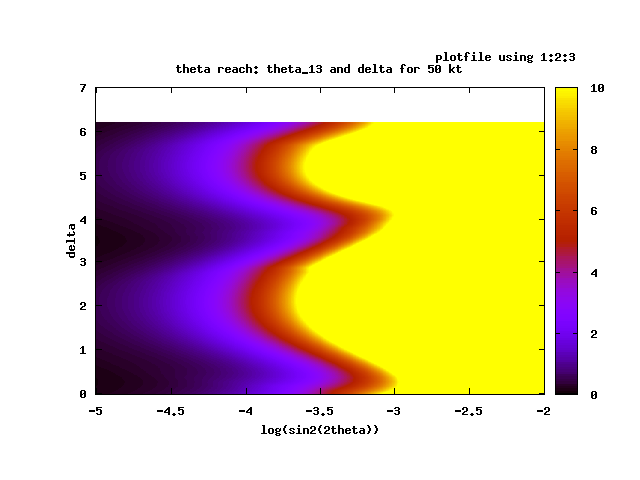}
\includegraphics[angle=0, width=7cm]{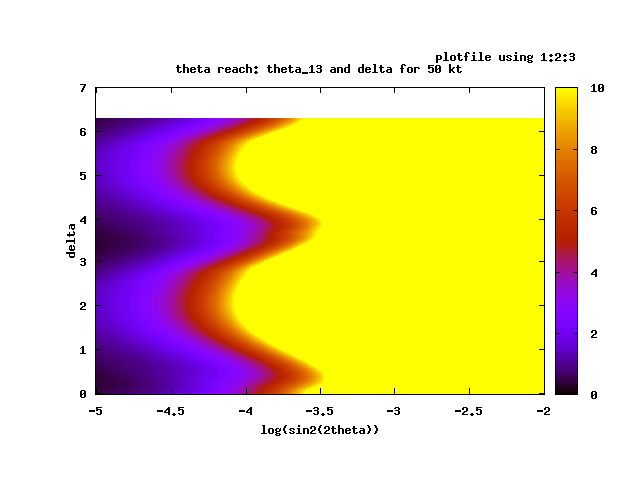}
\caption{\sf The reach of $\theta_{13}$ for a 50 kton detector and for a 150 kton detector with a neutrino factory and a wide band beam together for the same baseline.}
\label{theta_bw}
\end{center}
\end{figure}

\begin{figure}[tbhp]
\begin{center}
\includegraphics[angle=0, width=7cm]{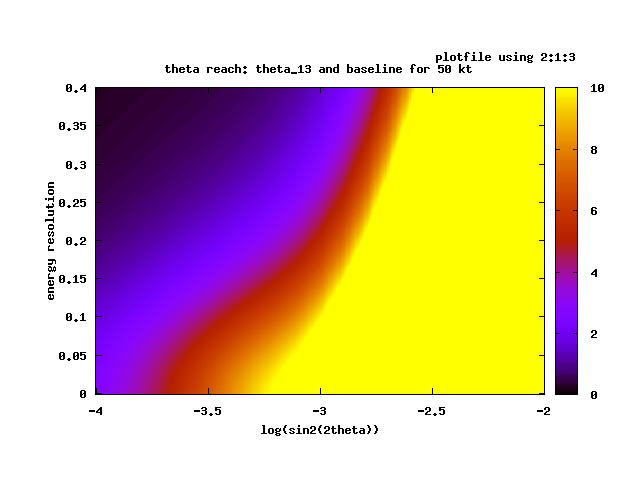}
\includegraphics[angle=0, width=7cm]{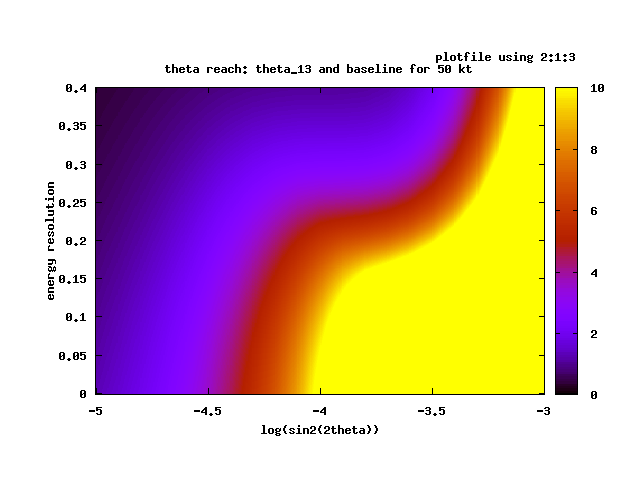}
\caption{\sf The capacity to measure the mass hierarchy for shown relative energy resolutions (150 kton detector for neutrino factory only), left $\delta=0.3$ and right $\delta=2.2$. The assumed 5 \% resolution is rather well-suited, as better resolution will not improve the capacity very much.}
\label{eres}
\end{center}
\end{figure}

To set the goals for the detector design I studied the performance for different detector properties. In figure \ref{eres} is shown as example the capacity to measure the mass hierarchy with different relative energy resolutions. That shows that a 20 \% resolution is required in order not to spoil the capacity to measure the mass hierarchy, and 5 \% resolution is sufficient.

\begin{figure}[tbhp]
\begin{center}
\includegraphics[angle=0, width=7cm]{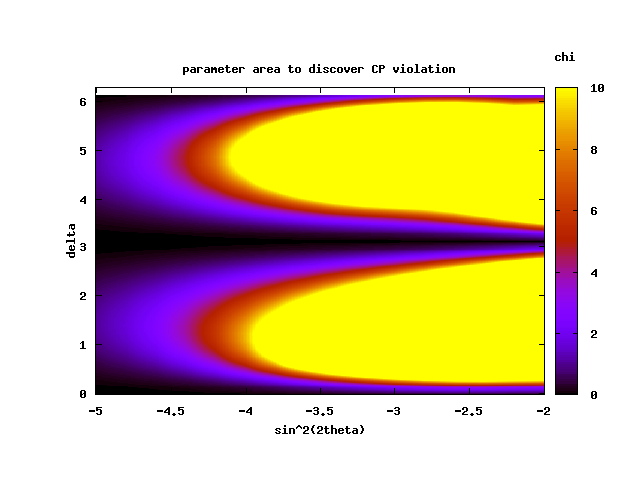}
\includegraphics[angle=0, width=7cm]{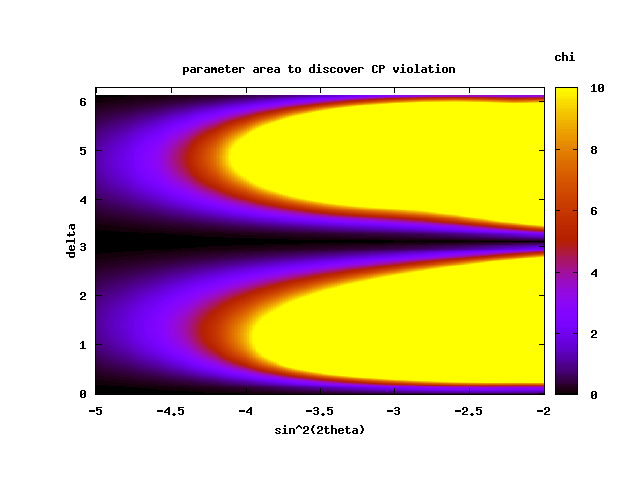}
\includegraphics[angle=0, width=7cm]{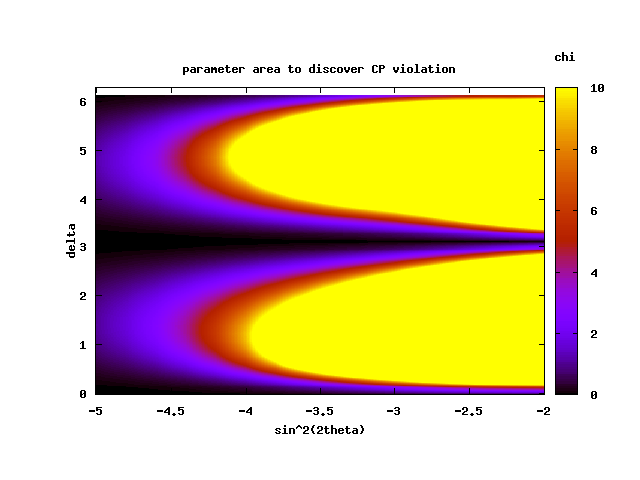}
\includegraphics[angle=0, width=7cm]{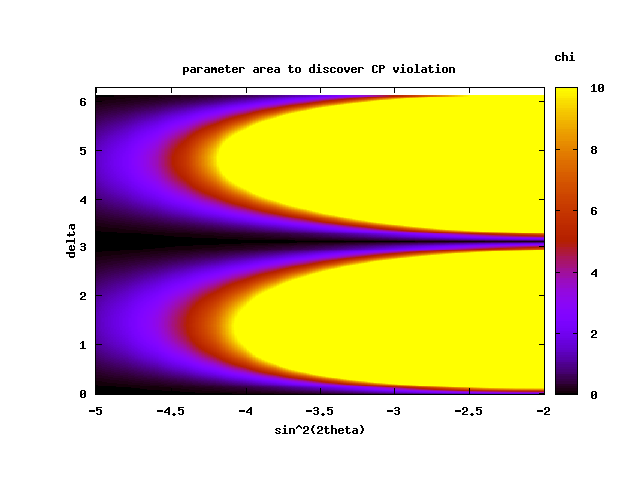}
\caption{\sf The capacity to study the CP violation with the neutrino factory only, including the golden and the platinum channel, with 150 kton. The figures are for 50 \%, 70 \%, 90 \% and 99 \% charge resolution (the first is no resolution at all). Evidently the statistical charge resolution brings little improvement.}
\label{platinum}
\end{center}
\end{figure}

I also studied the inclusion of the platinum channel with statistical cahrge recognition. As an example the results for CP violation are shown in Fig.~\ref{platinum}. Runs with 50 \% (pessimistic) 70 \% (realistic), 90 \% (optimistic) and 99 \% (utopic) charge identification probabilities were made, showing an improved performance at larger $\theta_{13}$ with a good electron charge recognition. A plain statistical charge identification is of little benefit.

\section{Conclusions}

Evidently a magnetized large volume liquid scintillator can be used as a far detector for the neutrino factory. For a low-energy (max. 5 GeV) beam a magnetized LENA with magnetic-field-tolerant photosensors would be suitable, higher energies need a dedicated detector aligned in the beam direction. While the 50 kton detector performes well, the vertical orientation being a small burden, a larger detector would be more optimal. The large volume scintillator is probably cheaper than a segmented scintillator and much more accurate than an iron calorimeter and has a rich research programme of its own in particle astronomy and nucleon decay searches. 

To reach the required accuracy an appropriate photodetection system working under magnetic field must be designed. Max a few-ns time resolution and multi-photon capacity are needed. Additional studies are needed to prove whether this is realistic. Further simulations are also required to settle the optimum magnetic field, initial studies suggesting that 0.1 T may be enough.

The low-energy neutrino factory with a 2288 km baseline was found to have a good capacity to study the neutrino properties. Neutrino oscillations up to $\sin^2 2\theta_{13} \sim 10^{-3} \dots 10^{-4}$ can be studied. For the lowest $\theta_{13}$ this beam outperforms the beta beam and any superbeam for a similar baseline. For a fair comparison, however, one has to take into account that at the cost of the neutrino factory one could build a whole lot of detectors for a conventional beam.

{\bf Acknowledgement}\\
This research was supported by the DFG cluster of excellence "Origin and Structure of the Universe". I thank LENA collaboration for inspiring co-operation, advice and help, E15 group of Franz von Feilitzsch for hospitality during my stay in Munich and LAGUNA consortium.

%\bibliographystyle{../h-elsevier}

%\bibliography{../cp}

\begin{figure}[tbhp]
\begin{center}
\includegraphics[angle=0, width=7cm]{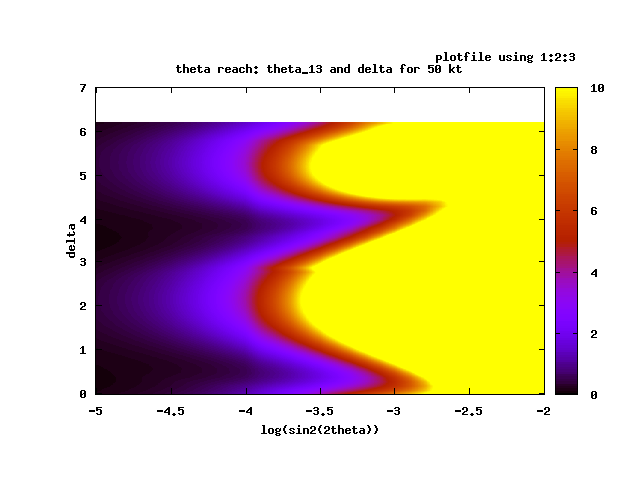}
\includegraphics[angle=0, width=7cm]{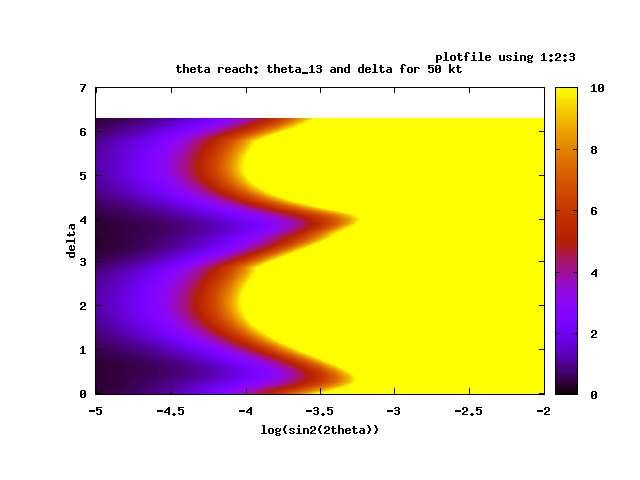}
\caption{\sf For comparison, the reach in $\theta_{13}$ with only a neutrino factory, 50 and 150 kton detectors.}
\label{theta_n}
\end{center}
\end{figure}

\begin{figure}[tbhp]
\begin{center}
\includegraphics[angle=0, width=7cm]{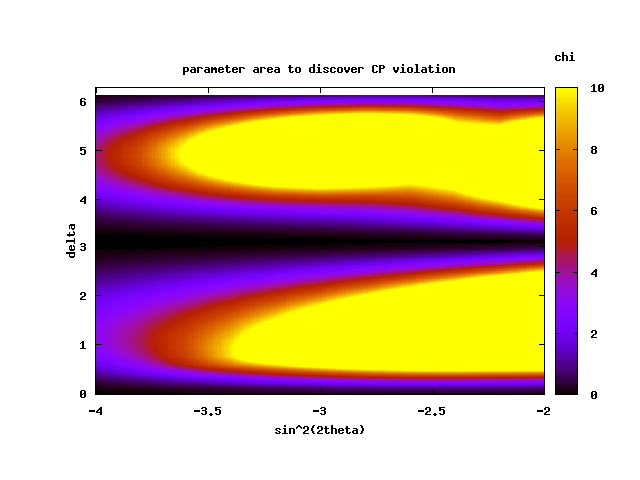}
\includegraphics[angle=0, width=7cm]{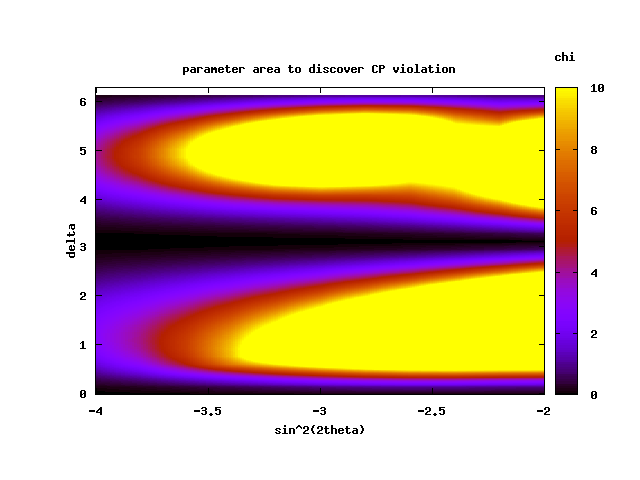}
\caption{\sf For comparison, the CP range with only a neutrino factory with a 50 kton detector. Left for the horizontal layout (no high-energy cut) and right for a vertical layout, i.e. reduced high-energy efficiency. The difference between these is barely visible, but the shapes differ remarkably from Fig.~\ref{cp_nw} that includes the conventional beam, too. For small $\theta_{13}$ and small CP violation the adjoint conventional beam is very important.}
\label{cp_n}
\end{center}
\end{figure}

\end{document}